\documentclass[12pt]{article}
\usepackage{amsmath, amssymb, amsthm}
\usepackage{graphicx,threeparttable,caption}
\usepackage[colorlinks=true, allcolors=blue]{hyperref}
\usepackage{mathrsfs}
\usepackage{natbib}
\usepackage{multirow}
\usepackage{float}
\usepackage{svg}
\usepackage{url}
\usepackage{enumerate}
\usepackage{chngcntr} 
\counterwithin{equation}{section}
\usepackage{color}
\usepackage{booktabs}
\usepackage{subfig} 
\usepackage{overpic} 
\usepackage{dsfont}
\usepackage{enumitem}
\usepackage{bm}
\usepackage[title]{appendix}
\usepackage{algorithm}
\usepackage{algpseudocode}
\allowdisplaybreaks[4]
\def\T{{ \mathrm{\scriptscriptstyle T} }}
\DeclareMathOperator*{\argmin}{argmin}

\newcommand{\E}{\mathbb{E}}
\newcommand{\ind}{\mathds{1}}

\newcommand{\lb}{\left(} 
\newcommand{\rb}{\right)} 
\newcommand{\br}[1]{\lb #1 \rb} 
\DeclareMathOperator{\cov}{cov}
\newtheorem{lemma}{Lemma}

\newtheorem{definition}{Definition}
\newtheorem{theorem}{Theorem}

\newtheorem{corollary}{Corollary}

\newtheorem{example}{Example}
\usepackage{authblk}
\usepackage{hyperref}
\newcommand{\blind}{1}

\begin{document}
\date{} 
\def\spacingset#1{\renewcommand{\baselinestretch}{#1}\small\normalsize}\spacingset{1}

\if1\blind
{
  \title{\bf Varying-Coefficient Fr\'echet Regression}
  \author[a]{Yanzhao Wang$^\dagger$}
  \author[b]{Jianqiang Zhang$^\dagger$}
  \author[a]{Wangli Xu$^*$}
  \affil[a]{Center for Applied Statistics and School of Statistics,
  Renmin University of China}
  \affil[b]{School of Statistics and Mathematics, Zhejiang Gongshang University}
  \maketitle
  \vspace{-1cm}
  \def\thefootnote{\fnsymbol{footnote}}
  \footnotetext[2]{These authors contributed equally to this work.}
  \footnotetext[1]{Corresponding author. Email: wlxu@ruc.edu.cn}
} \fi
\if0\blind
{
  \bigskip
  \bigskip
  \bigskip
  \begin{center}
    {\LARGE\bf Varying-Coefficient Fr\'echet Regression}
\end{center}
  \medskip
} \fi

\begin{abstract}
As a growing number of problems involve variables that are random objects, the development of models for such data has become increasingly important. This paper introduces a novel varying-coefficient Fr\'echet regression model that extends the classical varying-coefficient framework to accommodate random objects as responses. The proposed model provides a unified methodology for analyzing both Euclidean and non-Euclidean response variables. 
We develop a comprehensive estimation procedure that accommodates diverse predictor settings. Specifically, the model allows the  effect-modifier variable $U$ to be either Euclidean or non-Euclidean, while the predictors $\bm{X}$ are assumed to be Euclidean. Tailored estimation methods are provided for each scenario. To examine the asymptotic properties of the estimators, we introduce a smoothed version of the model and establish convergence rates through separate theoretical analyses of the bias and stochastic terms. The effectiveness and practical utility of the proposed methodology are demonstrated through extensive simulation studies and a real-data application.
\end{abstract}
\vfill

\newpage
\spacingset{1.9}
\section{Introduction}

Statistical analysis of non-Euclidean data has gained significant attention due to the increasing availability of complex data structures. Relevant examples include probability distributions, covariance matrices, phylogenetic trees, and graph networks. For instance, \citet{stock} studied the distribution of stock market returns to monitor structural changes over time. \citet{Iancovmat} analyzed brain images as covariance matrices to examine the diffusion of water molecules in the brain. \citet{Nye2016PrincipalCA} explored genomic data by investigating phylogenetic trees for dimensionality reduction and structure discovery. \citet{DubeyModelingtimevaryingrandom2022} examined the dynamics of time-varying networks, focusing on the evolving connectivity within brain networks and other complex systems. Given these developments, investigating the relationship between predictors and responses becomes crucial when the responses are metric space-valued  random objects rather than Euclidean variables.

In the literature, various models have been developed to analyze cases where the response consists of random objects while the predictors are Euclidean variables. Among these, \citet{PetersenFrechetregressionrandom2019} introduced the Fr\'echet regression method, which accommodates both linear and nonparametric relationships between the response and predictor variables. Building upon this foundation, \citet{BhattacharjeeSingleindexfrechet2023} proposed the single index Fr\'echet regression, which enables effective dimension reduction by projecting a multivariate predictor onto a single direction vector. \citet{TuckerPartiallyglobalfrechetregression2025} developed the partially-global Fr\'echet regression to combine the local and global Fr\'echet regression, which generalizes the partially linear regression model. This approach gives greater flexibility than purely global methods and improves accuracy compared to local methods. To mitigate the curse of dimensionality, \citet{QiuRandomforestweighted2024} introduced a random forest weighted local Fr\'echet regression, which relies on a locally adaptive kernel generated by random forests. More recently, \citet{Iao17072025} employed deep neural networks to develop deep Fr\'echet regression, which captures relationships between non-Euclidean responses and high-dimensional Euclidean predictors without imposing parametric assumptions. In summary, the development of Fr\'echet regression has progressed from parametric models towards flexible nonparametric and semiparametric frameworks, and further into high-dimensional and machine learning domains.

Although the existing literature has extensively studied various Fr\'echet regression models, these models naturally reduce to their Euclidean counterparts when the response variables lie in Euclidean space. For instance, global Fr\'echet regression simplifies to standard multiple linear regression \citep{PetersenFrechetregressionrandom2019}, while partially-global Fr\'echet regression reduces to classical partially linear regression model in the special case where random objects lie in Euclidean space \citep{TuckerPartiallyglobalfrechetregression2025}. Nevertheless, when dealing with Euclidean response variables, varying-coefficient models represent another important class of regression approaches that has not been similarly generalized to the Fr\'echet regression. These models capture dynamic relationships by allowing regression parameters to vary with covariates or over time, rather than being fixed constants. For example, \citet{varymodelpei2022} analyzed the relationship between income and pollution, showing that the impacts varied with energy consumption. Since their initial proposal by \citet{HastieVaryingcoefficientmodels1993}, these models have generated a substantial body of follow-up research. On the estimation side, kernel based local polynomial methods have been widely adopted for estimating functional coefficients by \citet{FanStatisticalestimationvarying1999}.  With further developments on efficient estimation, \citet{HuangVaryingcoefficientmodelsbasis2002} introduced basis function and spline based techniques to provide flexible global approximations suited for the repeated measurements. For inference on coefficient  variability, existing tests are based on the discrepancy between the restricted and unrestricted sum of squared residuals using smooth estimates of the varying coefficients. \citet{FanGeneralizedlikelihoodratio2001} developed the generalized likelihood ratio tests to examine whether coefficient functions are constant or vary over the domain, with desirable wilks phenomenon properties facilitating inference. In addition, \citet{HuangVaryingcoefficientmodelsbasis2002} proposed resampling-based subject bootstrap procedures to construct confidence regions and to perform hypothesis testing.

Varying-coefficient models represent a vital class of regression methods for characterizing relationships among variables in Euclidean spaces. Currently, no existing models have been developed for cases where the response variable is a random object. This study establishes the first estimation framework for Fr\'echet varying coefficient models, with the following principal innovations:

\begin{itemize}

\item [1.] We propose a novel varying-coefficient Fr\'echet regression model that extends the classical varying-coefficient framework to accommodate random objects as responses. This extension provides a unified methodology for analyzing response variables that are either Euclidean or non-Euclidean, without relying on parametric assumptions.

\item [2.] We develop an estimation procedure for the proposed varying-coefficient Fréchet regression model. Unlike existing methods for random object responses, which are typically limited to Euclidean predictors, our approach accommodates predictors that are themselves random objects residing in metric spaces.

\item [3.] To investigate the asymptotic convergence rate of the estimators, we introduce a smoothed version of the varying-coefficient Fr\'echet regression model. By separately analyzing the bias term and the stochastic term, we theoretically establish the convergence rate of the estimator. The effectiveness of the proposed method is validated through simulation studies and a real-data application.
\end{itemize}

The remainder of this paper is organized as follows. Section~\ref{sec:vcr} provides a comprehensive review of varying-coefficient models and corresponding estimation methods. Section~\ref{sec:vfr} introduces the proposed varying-coefficient Fr\'echet regression, including its definition and estimators. Section~\ref{sec:theorem} establishes the consistency and convergence rates of the proposed estimators. Section~\ref{sec:simulate} presents simulation studies with responses including scalars, probability distributions, and symmetric positive-definite matrices. Section~\ref{sec:apply} illustrates the empirical utility of the method through an application to human mortality data. Finally, Section~\ref{sec:discuss} concludes with a discussion of methodological implications and potential directions for future research.

\section{Varying-coefficient model}\label{sec:vcr}

To better introduce the varying-coefficient Fr\'echet model, this section first presents the properties of varying-coefficient models in Euclidean space. A varying-coefficient model with Euclidean responses takes the form
\begin{eqnarray} \label{var_model}
Y=\beta_0(U) + \bm{X}^\T \bm{\beta}(U)+\epsilon,
\end{eqnarray}
where $Y$ is  the response variable, the functional coefficient intercept $\beta_0(U)$ varies with the scalar $U\in \mathbb{R}$,  $\bm{X} = (X_1, X_2, \ldots, X_p)^\T\in\mathbb{R}^p$ is a $p$-dimensional vector of predictor variable, $\bm{\beta}(U) = (\beta_1(U), \beta_2(U), \ldots, \beta_p(U))^\T$ is a vector of coefficient function, and random error $\epsilon$ satisfying $\E(\epsilon|\bm{X},U)=0$. For every $u\in\mathbb{R}$, we have
\begin{eqnarray}\label{model_c}
\E(Y\mid U=u)=\beta_0(u)+\E(\bm X\mid U=u)^\T\bm{\beta}(u).   
\end{eqnarray}
From equations (\ref{var_model}) and (\ref{model_c}), we obtain
\begin{eqnarray*}\label{model_n}
Y - \mathbb{E}(Y \mid U=u) = (\bm{X} - \mathbb{E}(\bm{X} \mid U=u))^\T \bm{\beta}(u) + \epsilon.
\end{eqnarray*}
Thus,  $\bm{\beta}(u)$ can be obtained as follows
\begin{eqnarray}\label{beta}
{\bm{\beta}}(u) &= &\left( {\mathbb{E}}( (\bm{X} - {\mathbb{E}}(\bm{X}\mid U=u)) (\bm{X} - {\mathbb{E}}(\bm{X}\mid U=u))^\T\mid U=u) \right)^{-1} \nonumber \\
&& \times {\mathbb{E}}\left( (\bm{X} - {\mathbb{E}}(\bm{X} \mid U=u)) (Y - {\mathbb{E}}(Y \mid U=u))\mid U=u\right).
\end{eqnarray}

To estimate \( \bm{\beta}(u)\) in equation (\ref{beta}), we first estimate \( \mathbb{E}(\bm{X} \mid U=u) \) and \( \mathbb{E}(Y \mid U=u)\), denoted by \( \widehat{\mathbb{E}}(\bm{X} \mid U=u) \) and \( \widehat{\mathbb{E}}(Y \mid U=u) \), respectively. Suppose the random sample is $\{(Y_i, \bm{X}_i, U_i): i=1, \ldots, n\}$.
The estimators $\widehat{\mathbb{E}}(\bm{X}\mid U=u)$ and $\widehat{\mathbb{E}}(Y\mid U=u)$ are given by $\widehat{\mathbb{E}}(\bm{X}\mid U=u)=n^{-1}\sum_{i=1}^ns_{in}(u)\bm{X}_i$ and $\widehat{\mathbb{E}}(Y\mid U=u)=n^{-1}\sum_{i=1}^ns_{in}(u)Y_i$, where the weighting function $s_{in}(u)$ can be either the local constant weighting $s_{in}^{c}(u)$ \citep{Nadaraya1964,Watson1964} or the local linear weighting $s_{in}^{l}(u)$ \citep{FanLocalpolynomialmodelling1996} with their explicit forms given by
\begin{eqnarray}\label{kernel}
s_{in}^{c}(u)=\frac{K_h\left({U_i-u}\right)}{n^{-1}\sum_{j=1}^n K_h\left({U_j-u}\right)}, \quad s_{in}^{l}(u)=K_h(U_i-u) \frac{\widehat{\mu}_2(u)-\widehat{\mu}_1(u)(U_i-u)}{\widehat{\mu}_0(u) \widehat{\mu}_2(u)-\widehat{\mu}_1^2(u)}.
\end{eqnarray}
Here $K_h(\cdot)=h^{-1}K(\cdot/h)$, $K(\cdot)$ is a smoothing kernel function, $h$ is a bandwidth, and $\widehat{\mu}_j(u)=n^{-1}\sum_{i=1}^nK_h(U_i-u)(U_i-u)^j$ for $j=0, 1, 2$. 
Hereafter, unless otherwise specified, the weight function $s_{in}(u)$ may denote either the local constant weight $s_{in}^{c}(u)$ or the local linear weight $s_{in}^{l}(u)$, and will be abbreviated as $s_{in}$ for simplicity.
Using these notations, the estimator $\widehat{\bm{\beta}}(u)$ takes the form
\begin{eqnarray} \label{Hat_B}
\widehat{\bm{\beta}}(u)&=&
(\frac{1}{n}\sum_{i=1}^n s_{in} ( \bm{X}_i -\frac{1}{n}\sum_{j=1}^n s_{jn}\bm{X}_j) ( \bm{X}_i - \frac{1}{n}\sum_{j=1}^n s_{jn}\bm{X}_j )^\T )^{-1} \nonumber \\
&& \times \frac{1}{n}\sum_{i=1}^n s_{in}( \bm{X}_i - \frac{1}{n}\sum_{j=1}^n s_{jn}\bm{X}_j) ( Y_i - \frac{1}{n}\sum_{j=1}^n s_{jn}Y_j).
\end{eqnarray}
Based on equation (\ref{model_c}), the intercept term \(\beta_0(u)\) can be estimated by
\begin{equation} \label{inter_est}
\widehat{\beta}_0(u) = \widehat{\mathbb{E}}(Y \mid U=u) - \widehat{\mathbb{E}}(\bm{X} \mid U=u)^\T \widehat{\bm{\beta}}(u).
\end{equation}

Based on the estimated coefficients $\widehat{\bm{\beta}}(u)$ and $\widehat{\beta}_0(u)$, the predicted response at any given covariate vector 
 \(\bm{x}\)
 and location 
\(u\)
 is given by
\begin{eqnarray}\label{var_pre} 
\widehat{Y} =\widehat{\beta}_0(u) + \bm{x}^\T \widehat{\bm{\beta}}(u) =\frac{1}{n}\sum_{j=1}^n s_{jn} Y_j + (\bm{x} - \frac{1}{n}\sum_{k=1}^n s_{kn} \bm{X}_k)^\T \widehat{\bm{\beta}}(u)=\sum_{j=1}^n  c_j(\bm{x}, u) Y_j,
\end{eqnarray}
where $$
c_j(\bm{x}, u) = \frac{s_{jn}}{n} + (\bm{x} - \frac{1}{n}\sum_{k=1}^n s_{kn} \bm{X}_k)^\T \bm{\widehat{\Sigma}_{\bm{X}\mid u}}^{-1}  \sum_{i=1}^n\frac{s_{in}}{n}( \bm{X}_i - \frac{1}{n}\sum_{k=1}^n s_{kn}\bm{X}_k) (\delta_{ij} - \frac{s_{jn}}{n}),$$ 
with the conditional covariance matrix estimator defined as
$$\bm{\widehat{\Sigma}_{\bm{X}\mid u}} = \frac{1}{n}\sum_{i=1}^n s_{in}( \bm{X}_i - \frac{1}{n}\sum_{k=1}^n s_{kn} \bm{X}_k) ( \bm{X}_i - \frac{1}{n}\sum_{k=1}^n s_{kn} \bm{X}_k)^\T.$$
Here, the Kronecker delta $\delta_{i j}$ is defined by $\delta_{i j}=1$ if $i=j$ and $\delta_{i j}=0$ otherwise.
According to the above analysis and noting that $\sum_{j=1}^n c_j(\boldsymbol{x}, u)=1$, we can express the prediction in (\ref{var_pre}) as the solution to the following weighted least squares minimization problem
\begin{eqnarray}\label{var_pre1}
\widehat Y=\widehat{\beta}_0(u) + \bm{x}^\T \widehat{\bm{\beta}}(u)= \underset{y \in \mathcal{Y}}{\arg \min } \sum_{j=1}^n c_j(\bm{x}, u)\left(Y_j-y\right)^2.
\end{eqnarray}

\section{Varying-coefficient Fr\'echet regression}\label{sec:vfr} 
Before presenting the definition of varying-coefficient Fr\'echet regression and its estimation methods, we introduce some preliminary notation and assumptions. Specifically, let \((\mathcal{Y}, d)\) and $(\mathcal{U},\delta)$ be metric spaces, with metrics $d$ and $\delta$ defined on the sets $\mathcal{Y}$ and $\mathcal{U}$, respectively. We consider a random triple of objects \((\bm{X}, U, Y) \sim F\) defined in the product space \(\mathbb{R}^p \times \mathcal{U} \times \mathcal{Y}\). Here, \(\bm{X} = (X_1, X_2, \ldots, X_p)^\T \in \mathbb{R}^p\) represents a \(p\)-dimensional real-valued predictor and \(U \in \mathcal{U}\) is another predictor. The response variable \(Y \in \mathcal{Y}\) is a random object associated with both \(\bm{X}\) and \(U\). $F$ denotes the joint distribution of $(\bm{X}, U, Y)$, with marginal distributions $F_{\bm{X}}$, $F_U$ and $F_Y$. The conditional distributions $F_{Y \mid (\bm{X}, U)}$, $F_{\bm{X}\mid U}$ and $F_{Y \mid U}$ are assumed to exist. The conditional covariance matrix of $\bm X$ given $U=u$ is $\cov (\bm{X}\mid U=u)= \E ((\bm X-\E(\bm X\mid U=u))(\bm X-\E (\bm X\mid U=u))^\T\mid U=u)$. For a fixed $y\in\mathcal{Y}$, the conditional cross-covariance matrix between $\bm X$ and $d^2(Y,y)$ given $U=u$ is $\cov (\bm{X},d^2(Y,y)\mid U=u)= \E ((\bm X-\E(\bm X\mid U=u))(d^2(Y,y)-\E (d^2(Y,y)\mid U=u))\mid U=u)$.  

Building upon the above notation, we extend the varying-coefficient model \( Y = \beta_0(U) + \bm{X}^\top \bm{\beta}(U) + \epsilon \), where the response \( Y \) is Euclidean, to the case where \( Y \) is a random object. Motivated by \cite{PetersenFrechetregressionrandom2019}, which generalizes linear regression with Euclidean responses to settings where the response is a random object, we reformulate the varying-coefficient regression model by replacing the Euclidean distance with the intrinsic metric \( d \) on \( \mathcal{Y} \). This leads to the following definition of the varying-coefficient Fr\'echet regression model.

\begin{definition}[Varying-coefficient Fr\'echet regression]\label{VFR}
 We denote the conditional Fr\'echet regression function of $Y$ given $\bm{X}=\bm{x}$ and $U=u$ as
${m}_\oplus(\bm{x}, u) = \underset{y\in\mathcal{Y}}{\argmin} \, \E({d}^{2}({Y,y})\mid\bm{X}=\bm{x},U=u)$.
The varying-coefficient Fr\'echet regression model is said to hold if ${m}_{\oplus}\left({\bm{x}}, u\right)= s_{\oplus}\br{\bm{x},u}$ for any \(\bm{x}\in\mathbb{R}^p\) and \(u\in\mathcal{U}\), where $s_{\oplus}\br{\bm{x},u}$ is defined by
\begin{equation}\label{var_fre_model1}
s_{\oplus}\br{\bm{x},u}=\underset{y \in  \mathcal{Y}}{\arg \min } \, {S}_{ \oplus}\left( {y;\bm{x}, u}\right),
\end{equation}with
\begin{equation*}
{S}_{ \oplus  }\left( {y;\bm{x},u}\right)  = {w}_{0}\left( {y;u}\right)+{w}_{1}\left(\bm{x},u\right)w_2^{-1}\br{u}w_3\br{y;u}.
\end{equation*}
The terms \(w_0(y; u)\), \(w_1(\bm{x}, u)\), \(w_2(u)\), and \(w_3(y; u)\) in \(S_{\oplus}(y; \bm{x}, u)\) are defined as follows, ${w}_{0}( y;u) =\E(d^2\br{Y,y}\mid U=u)$, $w_1\br{\bm{x},u}=\bm{x}^\T-\E\br{\bm{X}\mid U=u}^\T$, $w_2\br{u}= \cov({\bm{X}\mid U=u})$, and ${w}_{3}(y;u)=
\cov (\bm{X},d^2\br{Y,y}\mid U=u)$.
\end{definition}

An important case arises when the random objects lie in a Hilbert space $\mathcal Y$ equipped with inner product $\langle \cdot,\cdot \rangle$ and induced norm $\|\cdot\|_{\mathcal Y}$. Under mild assumptions, the minimization problem \eqref{var_fre_model1} has an explicit solution in this setting, thanks to the linearity of the inner product and Riesz representation theorem. 
To facilitate a detailed exposition of this case, we introduce the following notation. For $p>1$, let $\mathcal Y^p$ denote the $p$-fold Cartesian product of $\mathcal Y$, equipped with inner product $\langle \bm y, \bm y' \rangle_p = \sum_{i=1}^p \langle  y_i,  y'_i \rangle$, $\bm y, \bm y' \in \mathcal Y^p$. Then $\mathcal Y^p$ is itself a Hilbert space. For a $p \times p$ matrix $A$, $\bm x \in \mathbb{R}^p$, $y\in \mathcal Y$ and $\bm y \in \mathcal Y^p$, we define $A\bm y\in \mathcal Y^p$ with elements $(A\bm y)_i = \sum_{j=1}^p A_{ij}  y_j$, $\bm x^\T \bm y = \sum_{i=1}^p  x_i y_i \in \mathcal Y$ and $\bm xy\in \mathcal Y^p$ with elements $(\bm xy)_i = x_iy.$  

\begin{lemma}[Varying–coefficient Fr\'echet regression in Hilbert space]\label{thm:hilbert-reduction}
For a given $u \in \mathcal{U}$, set $\mu(u):=\E(\bm X\mid U=u)\in\mathbb R^p$,  and assume $\E(\|Y\|_\mathcal Y^2 \mid U=u) <\infty$.
Define $s_u(\bm x,\bm X):=1+(\bm x-\mu(u))^\T w_2^{-1}(u)(\bm X-\mu(u))$ and thus $s_{\oplus}(\bm x,u)=\arg\min_{ y\in\mathcal Y}\E(s_u(\bm x,\bm X)\,\|Y- y\|_\mathcal Y^2\mid U=u)$. Then there exist unique elements $\gamma_0\in\mathcal Y$ and $\gamma_1\in \mathcal Y^p$, such that for all $y\in\mathcal{Y}$ and $\bm y\in\mathcal{Y}^p$, $\E(\langle Y, y\rangle\mid U=u)=\langle \gamma_0(u), y\rangle$ and $\E( \langle (X-\mu(u))Y, y\rangle_p\mid U=u)=\langle \gamma_{1}(u), y\rangle_p$. Define $\bm \beta(u):=w_2^{-1}(u)\gamma_1(u)$ and $\beta_0(u):=\gamma_0(u)-\mu(u)^\T w_2^{-1}(u)\gamma_1(u)$, the solution to \eqref{var_fre_model1} is $s_{\oplus}(\bm x,u)=\beta_0(u)+\bm x^\T \bm\beta(u)$.
\end{lemma}


For responses that are random objects in a Hilbert space, Lemma \ref{thm:hilbert-reduction} provides explicit solutions to the minimization problems \eqref{var_fre_model1} that define the varying-coefficient Fr\'echet regression. In particular, when $\mathcal Y$ is Euclidean space, the varying-coefficient Fr\'echet regression model can simplify to the classical varying-coefficient model (\ref{var_model}).

To estimate the varying-coefficient Fr\'echet regression model, it is necessary to estimate $S_{\oplus}(y; \bm{x}, u)$, which is denoted as $\widehat{S}_{\oplus}(y; \bm{x}, u)$, and correspondingly obtain the estimate of $s_{\oplus}(\bm{x}, u)$, denoted as $\widehat{s}_{\oplus}(\bm{x}, u)$. Based on an independent and identically distributed sample $\{(\bm X_i,U_i,Y_i):i=1,2,\ldots n\}$ from the joint distribution $F$, the estimate for the varying-coefficient Fr\'echet regression model is defined as
\begin{equation} \label{var_fre_est1} 
{\widehat{s}}_{ \oplus  }\left( {\bm{x},u}\right)  = \underset{y \in  \mathcal{Y}}{\arg \min }\,{\widehat{S}}_{n}\left( {y;\bm{x},u}\right),  
\end{equation}
where
\begin{equation*}
{\widehat{S}}_{n}\left( {y;\bm{x},u}\right)  = {\widehat{w}}_{0}\left( y;u\right)+ {\widehat{w}}_{1}\left( \bm{x},u\right){\widehat{w}}_{2}^{-1}\br{u} \widehat{w}_{3}\left( y;u\right).
\end{equation*}
The estimators $\widehat{w}_0(y;u)$, $\widehat{w}_1(\bm{x}, u)$, $\widehat{w}_2(u)$, and $\widehat{w}_3(y;u)$ are the respective estimates of ${w}_0(y;u)$, ${w}_1(\bm{x}, u)$, ${w}_2(u)$, and ${w}_3(y;u)$. We now consider two cases based on the domain of the predictor $u$: Euclidean space and non-Euclidean space.

For the case where $u$ lies in a Euclidean space, the expressions for $\widehat{w}_0(y;u)$, $\widehat{w}_1(\bm{x}, u)$, $\widehat{w}_2(u)$, and $\widehat{w}_3(y;u)$ are given as follows
\begin{eqnarray} \label{w_est}
 &&\widehat{w}_0(y;u)= \frac{1}{n} \sum_{i=1}^n s_{in}(u) d^2(Y_i, y), \qquad \widehat{w}_1(\bm{x}, u)\;=\; \bm{x}^\T-\frac{1}{n}\sum_{i=1}^n s_{in}(u) \bm{X}_i^\T, \nonumber \\
&&\widehat{w}_2\br{u}=\frac{1}{n}\sum_{i=1}^ns_{in}(u)(\bm{X}_i-\frac{1}{n} \sum_{j=1}^n s_{jn}\left(u\right) \bm{X}_j)(\bm{X}_i-\frac{1}{n}\sum_{j=1}^n s_{jn}\left(u\right) \bm{X}_j)^\T,\\
&&{\widehat{w}}_{3}( {y;u}) =\frac{1}{n}\sum_{i=1}^ns_{in}(u)({{\bm{X}}_{i} - \frac{1}{n}\sum_{j=1}^n{s}_{jn}( u) {\bm{X}}_{j}})({{d}^{2}( {{Y}_{i},y})  - \frac{1}{n}\sum_{j=1}^n{s}_{jn}(u) {d}^{2}( {{Y}_{j},y})}). \nonumber
\end{eqnarray}
Here the weight function $s_{in}(u)$ may denote either the local constant weight $s_{in}^c(u)$ or the local linear weight $s_{in}^l(u)$ in (\ref{kernel}). Corresponding to the varying-coefficient model with a Euclidean response variable, we have
\begin{equation*}
\widehat{s}_{\oplus}(\bm{x}, u) = \arg\min_{y \in \mathcal{Y}} \sum_{j=1}^n c_j(\bm{x}, u) d^2(Y_j, y),
\end{equation*}
where $c_j(\bm{x}, u)$ is defined in (\ref{var_pre1}).

For the case where $u$ is a random object, the expressions for $\widehat{w}_0(y;u)$, $\widehat{w}_1(\bm{x}, u)$, $\widehat{w}_2(u)$, and $\widehat{w}_3(y;u)$ are the same as those in (\ref{w_est}), except for the weight function $s_{in}(u)$, which takes the following form
\begin{eqnarray}\label{sr}
s_{in}^{r}(u)=\frac{K_h(\delta(U_i,u))}{n^{-1}\sum\nolimits_{j=1}^n K_h\left(\delta(U_j,u)\right)},
\end{eqnarray}
where $\delta$ denotes the metric on the space $\mathcal{U}$.

\section{Theoretical properties}\label{sec:theorem}

In this section, we establish the theoretical properties of the estimators for varying-coefficient Fr\'echet regression under Euclidean and non-Euclidean settings of the predictor $U$. For the following theoretical result, we assume that $\bm X$ has bounded support and $\mathcal Y$ is a totally bounded metric space. To derive the convergence rate of the distance $d({{s}_{\oplus}(\bm{x},u),{\widehat{s}}_{\oplus}(\bm{x},u)})$, we first introduce the smoothed version of the varying-coefficient Fr\'echet regression ${{s}}_{ \oplus}\br{\bm{x},u}$ as
\begin{equation}\label{var_fre_pop}
{\widetilde{s}}_{ \oplus  }\left({\bm{x},u}\right) = \underset{y \in  \mathcal{Y}}{\arg\min }\, {\widetilde{S}}_{n}\left( {y;\bm{x},u}\right),
\end{equation}
where
\begin{equation*}
{\widetilde{S}}_{n}\left( {y;\bm{x},u}\right)= {\widetilde{w}}_{0}\br{y;u}+{\widetilde{w}}_{1}\br{\bm{x},u}{\widetilde{w}}_{2}^{-1}\left( u \right) {\widetilde{w}}_{3}\br{y;u}.
\end{equation*}
The terms \(\widetilde{w}_0(y; u)\), \(\widetilde{w}_1(\bm{x}, u)\), \(\widetilde{w}_2(u)\), and \(\widetilde{w}_3(y; u)\) are defined by the following expressions,
\begin{eqnarray*} 
&&{\widetilde{w}}_{0}\left( {y;u}\right) = \E({\zeta }_{h}\br{{U,u}}d^2\br{Y,y}),\qquad{\widetilde{w}}_{1}\br{\bm{x},u} = \bm{x}^\T-\E\br{{\zeta }_{h}\br{{U,u}}\bm{X}}^\T, \\
&&{\widetilde{w}}_{2}\br{u} = \E({\zeta}_{h}({U,u})(\bm{X}-\E({\zeta}_{h}({U,u})\bm{X}))(\bm{X}-\E({\zeta}_{h}({U,u})\bm{X}))^\T),\\
&&{\widetilde{w}}_{3}\left( y;u\right)= \E({\zeta}_{h}({U,u})(\bm{X}-\E({\zeta}_{h}({U,u})\bm{X}))(d^2({Y,y})-\E({\zeta }_{h}({U,u})d^2({Y,y})))).
\end{eqnarray*}
The form of weight function $\zeta_h(U, u)$ depends on the space of the predictor $U$. For Euclidean $U$, the weight function may denote either $\zeta_h^l(U, u)$ or $\zeta_h^c(U, u)$, given respectively by
\begin{eqnarray}\label{zeta}
\zeta_h^l(U, u)=K_h(U-u) \frac{\widetilde{\mu}_2(u)-\widetilde{\mu}_1(u)(U-u)}{\widetilde{\sigma}_0^2(u)},\quad \zeta _h^c({U,u}) =\frac{K_h(U-u)}{\E\br{K_h(U-u)}},    
\end{eqnarray}
where $\widetilde{\mu}_j(u)=\E\left(K_h(U-u)(U-u)^j\right)$ for $j=0,1,2$, and $\widetilde{\sigma}_0^2(u)=\widetilde{\mu}_0(u) \widetilde{\mu}_2(u)-\widetilde{\mu}_1^2(u)$.
For non-Euclidean $U$, the weight function is denoted by $\zeta_h^r(U, u)$, which is defined as
\begin{eqnarray}\label{zeta_Non}
\zeta _h^r({U,u}) =\frac{K_h\br{\delta\br{U,u}}}{\E\br{K_h\br{\delta\br{U,u}}}}.    
\end{eqnarray}

\subsection{Predictor \texorpdfstring{$U$}{U} in Euclidean space}

In this subsection, we study the properties of the varying-coefficient Fr\'echet regression estimator with a Euclidean predictor $U$. We focus on points $\bm{x} \in \mathbb{R}^p$ and $u \in \mathbb{R}$ for which the marginal densities satisfy $f_{\bm{X}}(\bm{x}) > 0$ and $f_{U}(u) > 0$. The empirical estimator corresponding to the weight function $\zeta_h(U, u)$ in $\widetilde{s}_{\oplus}(\bm{x}, u)$ is $s_{in}(u)$ in $\widehat{s}_{\oplus}(\bm{x}, u)$. The weight function $\zeta_h(U, u)$ can be chosen as either $\zeta_h^l(U, u)$ or $\zeta_h^c(U, u)$, with the corresponding estimators denoted by $s_{in}^l(u)$ and $s_{in}^c(u)$, respectively. In this paper, our theoretical analysis focuses on the more complex case in which the weight function takes the form $\zeta_h^l(U, u)$.

With the notation established, we aim to obtain the convergence rate for the distance $d\left({{s}_{\oplus}\left( \bm{x},u\right) ,{\widehat{s}}_{\oplus}\left(\bm{x},u\right) }\right)$. The analysis is based on the bias–variance decomposition: we first derive the convergence rate of the bias term $d\left({{s}_{\oplus}\left( \bm{x},u\right),{\widetilde{s}}_{\oplus}\left(\bm{x},u\right) }\right)$, and then analyze the stochastic variation term $d\left({\widetilde{s}_{\oplus}\left(\bm{x},u\right),{\widehat{s}}_{\oplus}\left(\bm{x},u\right) }\right)$. The necessary assumptions (A1)--(A6) are stated below.\\
(A1) The object $s_{\oplus}(\bm{x}, u)$ exists and is unique. There exist a positive integer $n_0$ such that for all $n \geq n_0$, $\widetilde{s}_{\oplus}(\bm{x}, u)$ and $\widehat{s}_{\oplus}(\bm{x}, u)$ exist and are unique, the latter almost surely. Additionally, for any $\epsilon>0$,
\begin{align*}
  \inf _{d\left(y, s_{\oplus}(\bm{x}, u)\right)>\epsilon}\big\{S_{\oplus}(y;\bm{x}, u)-S_{\oplus}\left(s_{\oplus}(\bm{x},u);\bm{x}, u\right)\big\}>0,\\
\mbox{and} \quad \liminf _{n \rightarrow \infty} \inf _{d\left(y, \tilde{s}_{\oplus}(\bm{x}, u)\right)>\epsilon}\big\{\widetilde{S}_n(y ; \bm{x}, u)-\widetilde{S}_n\left(\widetilde{s}_{\oplus}(\bm{x}, u) ; \bm{x}, u\right)\big\}>0.  
\end{align*}
(A2) Let $B_{\mathcal{Y}}\left(s_{\oplus}(\bm{x}, u), \delta\right) \subseteq \mathcal{Y}$ be the ball of radius $\delta$ centered at $s_{\oplus}(\bm{x}, u)$ and $N(\epsilon,$ $ B_{\mathcal{Y}}(s_{\oplus}(\bm{x}, u), \delta),d)$ be its covering number using balls of size $\epsilon$ with respect to the metric $d$. Then $\int_0^1 (1+\log N\left(\delta\epsilon, B_{\mathcal{Y}}\left(s_{\oplus}(\bm{x}, u), \delta\right), d\right))^{1/2} d \epsilon=O(1)$ as $\delta \rightarrow 0^+$.\\
(A3) The kernel $K(\cdot)$ is a probability density function, symmetric around zero. Furthermore, defining $K_{kj}=\int_{\mathbb{R}}K^k(u)u^jdu$, $K_{12}$, $K_{14}$, $K_{22}$, $K_{24}$ and $K_{26}$ are both finite.\\
(A4) The densities $f_U(\cdot)$, $f_{U|Y = y}(\cdot)$, $f_{U|{X}_i=x_i,X_j ={x}_j}(\cdot)$ and $f_{U|{X}_i={x}_i,Y=y}(\cdot)$ exist and are twice continuously differentiable. The $\sup_{y,u} |f_{U|Y = y}^{\prime\prime}(u)|$ and $\sup_{{x}_i,u} |f_{U|{X}_i = {x}_i}^{\prime\prime}(u)|$ are finite. For any open set $A\subseteq \mathcal{Y}$, $\int_Ad\mathbb{P}_{Y\mid U=u}$, $\int_Ad\mathbb{P}_{X_i\mid U=u}$, $\int_Ad\mathbb{P}_{X_i,X_j\mid U=u}$ and $\int_Ad\mathbb{P}_{Y,X_j\mid U=u}$ is continuous as a function of $u$, with $i,j=1,2,\ldots,p$.\\  
(A5) There exist $\eta_1>0, C_1>0$, and $\beta_1>1$ such that whenever $d\left(y, s_{\oplus}(\bm{x}, u)\right)<\eta_1$, 
$$
S_{\oplus}(y ; \bm{x}, u)-S_{\oplus}\left(s_{\oplus}(\bm{x}, u) ; \bm{x}, u\right) \geq C_1 d\left(y, s_{\oplus}(\bm{x}, u)\right)^{\beta_1}.
$$
(A6) There exist $\eta_2>0, C_2>0$, and $\beta_2>1$ such that whenever $d\left(y, \widetilde{s}_{\oplus}(\bm{x}, u)\right)<\eta_2$, 
$$
\liminf _{n\rightarrow \infty}\big\{\widetilde{S}_{n}(y ; \bm{x}, u)-\widetilde{S}_{n}\left(\widetilde{s}_{\oplus}(\bm{x}, u) ; \bm{x}, u\right)\big\} \geq C_2 d\left(y, \widetilde{s}_{\oplus}(\bm{x}, u)\right)^{\beta_2}.
$$

Assumption (A1) forms the foundational basis for establishing the consistency of $M$-estimator ${s}_{\oplus}(\bm{x}, u)$ (see, e.g., Corollary 3.2.3 of \cite{VanDerVaartWeakconvergenceempirical1996}), as it guarantees that the weak convergence of the empirical process $\widehat{S}_{\oplus}(\bm{x},u)$ to the population process $\widetilde{S}_{\oplus}(\bm{x}, u)$ in turn implies convergence of their minimizers. Furthermore, existence follows immediately if $\mathcal{Y}$ is compact. Building upon this, Assumption (A2) and (A5) impose covering number requirements and curvature constraints, which originate from empirical process theory and jointly regulate the local behavior of the deviation $\widehat{S}_{\oplus}(\bm{x},u)-\widetilde{S}_{\oplus}(\bm{x},u)$ near the optimum to establish convergence rates. Meanwhile, the assumptions on the kernel function in (A3) and the conditional density specifications in (A4) align with standard prerequisites in nonparametric local regression frameworks. Assumption (A6) is  an extension of Assumption (A5), which controls the convergence rate of $M$-estimator $\widetilde{s}_{\oplus}(\bm{x}, u)$. Notably, in the Euclidean space  $\mathcal{Y}=\mathbb{R}$ endowed with the Euclidean distance $d$, Assumptions (A5) and (A6) hold with $\beta_1=\beta_2=2$. 

\begin{theorem}\label{linear_theorem1}
Under Assumptions (A1)-(A5), the bias component satisfies
\begin{eqnarray*}
 d\left( {{s}_{\oplus}\left(\bm{x},u\right) ,{\widetilde{s}}_{\oplus}\left(\bm{x},u\right) }\right)  = O({h}^{2/\left( {{\beta}_{1} - 1}\right) }),   
\end{eqnarray*}
as the bandwidth $h\rightarrow0$, where the ${\beta}_{1}$ is specified in Assumption (A5).
\end{theorem}
This result quantifies the convergence rate of the bias, which depends on the local geometric structure of the metric space around the minimizer through $\beta_1$. The parameter $\beta_1$ characterizes the order of smoothness of $S_\oplus(y;\bm x,u)$ and thereby determines the bias decay rate $O({h}^{2/\left( {{\beta}_{1} - 1}\right)})$. In the Euclidean case, $\beta_1=2$ corresponds to the standard quadratic curvature condition, yielding the familiar bias rate of order $O(h^2)$ in classical nonparametric regression, which is known to be optimal.
\begin{theorem}\label{linear_theorem2}
Under Assumptions (A1)-(A3), and (A6), if $h\rightarrow0$ and  $nh\rightarrow\infty$ as $n\rightarrow\infty$, the stochastic component exhibits the probabilistic convergence rate
$$
d\left( {{\widetilde{s}}_{ \oplus  }\left(\bm{x},u\right),{\widehat{s}}_{ \oplus  }\left(\bm{x},u\right) }\right)  = {O}_{p}\big( {(nh) }^{{-1}/({2({{\beta }_{2} - 1})})}\big),
$$
where ${\beta}_{2}$ is specified in Assumption (A6).
\end{theorem}
Theorem \ref{linear_theorem2} characterizes the convergence rate of variation term $d\left( {{\widetilde{s}}_{ \oplus  }\left(\bm{x},u\right),{\widehat{s}}_{ \oplus  }\left(\bm{x},u\right) }\right)$. The convergence rate depends on $\beta_2$, which reflects the local curvature of the $\widetilde{S}_\oplus(y;\bm x,u)$ around its minimizer. This result is consistent with the convergence rate established for local Fr\'echet regression in \cite{PetersenFrechetregressionrandom2019}. In the Euclidean case, setting $\beta_2=2$ recovers the optimal variance rate $O_p((nh)^{-1/2})$.

\begin{corollary}\label{linear_corollary}

Under the assumptions of Theorem \ref{linear_theorem1} and Theorem \ref{linear_theorem2}, the distance between $s_{\oplus}(\bm{x},u)$ and $\widehat{s}_{\oplus}(\bm{x},u)$ satisfies the following rate
\begin{eqnarray*}
d\left(s_{\oplus}(\bm{x},u), \widehat{s}_{\oplus}(\bm{x},u)\right) =  O_p\left( h^{2/(\beta_1 - 1)}+ (nh)^{-1/(2(\beta_2 - 1))} \right).
\end{eqnarray*}
\end{corollary}

In general, the rate of convergence for the $d( {{s}_{ \oplus  }(\bm{x},u) ,{\widehat{s}}_{\oplus}(\bm{x},u) })$ is determined by the local geometry near the minimum as quantified in (A5) and (A6). The 
Corollary \ref{linear_corollary} can be obtained by applying the triangle inequality in metric space to combine Theorems \ref{linear_theorem1} and \ref{linear_theorem2}. This result  presents the  bias-variance trade-off analogous to that in classical nonparametric regression, and the optimal $h$ is of order $n^{-\frac{\beta_1-1}{4\beta_2+\beta_1-5}}$. Additionally, as $h\rightarrow0$ and  $nh\rightarrow\infty$, we obtain the result which matches the special case where $\mathcal{Y}=\mathbb{R}$ and $d$ is the Euclidean distance. Specifically, when $\beta_1=\beta_2=2$,  Corollary \ref{linear_corollary} guarantees that $d( {{s}_{ \oplus  }(\bm{x},u) ,{\widehat{s}}_{ \oplus}(\bm{x},u) })=O_p({h^2+({nh})^{-1/2}})$, which is the same as the result in \cite{CaiEfficientestimationinferences2000} for local varying-coefficient models with real valued responses.

\subsection{Predictor \texorpdfstring{$U$}{U} in non-Euclidean space}

In this subsection, we primarily investigate the properties of the varying-coefficient Fr\'echet regression estimator with a non-Euclidean predictor $U$. The weight function $\zeta_h(U, u)$ is given by $\zeta_h^r(U, u)$ in  (\ref{zeta_Non}). The corresponding empirical weight in $\widehat{s}_\oplus(\bm x,u)$ is $s_{i n}^r\br{u} = K_h\br{\delta\br{U_i,u}}/(n^{-1}\sum_{j=1}^n K_h\br{\delta\br{U_j,u}})$, as defined in  (\ref{sr}).

Because the probability of non-Euclidean $U$ depends on the metric space \(\left(\mathcal{U},\delta \right)\), we define the small ball probability of random object \(U \in  \mathcal{U}\) as ${\varphi }_{\mathcal{U},u}(h)  = \mathbb{P}( {U \in  {B}_{\mathcal{U}}( {u,h}) })$, where ${B}_{\mathcal{U}}\left( {u,h}\right)  = \{  {{u}^{\prime } \in \mathcal{U}:\delta \left( {{u}^{\prime },u}\right) \leq  h}\}$. With the small ball probability definitions, we require the following assumptions to handle the predictor $U$ in the metric space.\\
(A7) For any $\epsilon  > 0$, $\mathbb{P}\left( {U \in  {B}_{\mathcal{U}}\left( {u,\epsilon }\right) }\right)  = {\varphi }_{\mathcal{U},u}\left( \epsilon \right)  > 0$.\\
(A8)  As $n\rightarrow\infty$, we have $h\rightarrow0$, ${\log n}/{(n{\varphi }_{\mathcal{U},u}\left( h\right) }) \rightarrow 0$, and $nh\rightarrow\infty$.\\
(A9) There exist constants $0<c_1\le c_2<\infty$ and $C$, such that $c_1\ind_{[0,1]} (\cdot)\le K (\cdot)\le c_2\ind_{[0,1]}(\cdot)$ and $\int K_h(\delta(u', u)) \delta(u', u) d \nu(u') \leq C h \int K_h(\delta(u', u)) d \nu(u')$, where $\ind_{[0,1]}(\cdot)$ denotes the indicator function.\\
(A10) The conditional density $f_{U \mid Y=y}(u)$, $f_{U \mid X_i=x_i}(u)$, $f_{U \mid X_i=x_i,X_j=x_j}(u)$, $f_{U \mid X_i=x_i,Y=y}(u)$ for $i,j=1,2\ldots,p$, and the marginal density $ f_U(u) $ exist (with respect to a reference measure $\nu$) and are Lipschitz continuous. That is, there exists a constant $ C > 0 $ such that for all $ u' \in {B}_{\mathcal{U}}( {u,h})$ and  any conditional density $f(\cdot)$ in the above, $|f(u') - f(u)| \leq C \delta(u', u)$.

Assumption (A7) ensures that local neighborhood $B_{\mathcal{U}}(u,\epsilon)$ around point $u$ are not empty and provides sufficient sample probability mass for local smoothing methods such as kernel estimation. 
In Assumption (A8), as the sample size $n \to \infty$, the requirement $h \to 0$ eliminates asymptotic bias, the condition $\log n /(n \varphi_{\mathcal{U},u}(h)) \to 0$ regulates stochastic variability, and the growth condition $nh \to \infty$ guarantees that the effective number of local observations diverges.
Assumption (A9) imposes mild conditions on the kernel function $K(\cdot)$. It requires that $K(\cdot)$ is bounded between two positive constants within the unit support and vanishes outside, while also satisfying an integral inequality involving the distance function $\delta(\cdot,\cdot)$. In addition, it  guarantees that $
    \E[\zeta_h(U,u)|Y=y] = f_{U \mid Y=y}(u)f^{-1}_U(u)\left(1 + O(h)\right)$.
Assumption (A10) requires the existence and Lipschitz continuity of the conditional densities and the marginal density. The Lipschitz condition ensures that these densities vary smoothly with respect to $u$, which is crucial for controlling approximation errors in estimation and establishing convergence results. 

Next, we present the main theoretical results for the estimator when $U$ is treated as a random object. Similar to the case where $u$ is a Euclidean variable, we first analyze the bias term $d(s_{\oplus}(\bm{x},u),\widetilde{s}_{\oplus}(\bm{x},u))$, then derive the convergence rate of the stochastic term $d(\widetilde{s}_{\oplus}(\bm{x},u),\widehat{s}_{\oplus}(\bm{x},u))$. A corollary combining these results is provided to establish the overall convergence rate of $d(s_{\oplus}(\bm{x},u), \widehat{s}_{\oplus}(\bm{x},u))$.

\begin{theorem}\label{mix_theorem1}
Under Assumptions (A1), (A2), (A5), and (A8)-(A10),  the following holds
$$
d\left( {{s}_{ \oplus  }\left( {\bm{x},u}\right) ,{\widetilde{s}}_{ \oplus  }\left( {\bm{x},u}\right) }\right)  = O(h^{\frac{1}{\beta_1 - 1}}),
$$
where $\beta_1$ is  specified in Assumption (A5).
\end{theorem}
This result establishes the convergence rate of the bias term $d(s_{\oplus}(\bm{x},u),\widetilde{s}_{\oplus}(\bm{x},u))$. The rate is determined by the Lipschitz continuity in (A10) and the local geometry of the metric space through $\beta_{1}$ in (A5). When $\mathcal{Y}=\mathbb{R}$, we  have $\beta_1=2$, and the convergence rate becomes $O(h)$. One might naturally ask why the result here is $O(h)$, given that local constant smoothing with a symmetric kernel is well known to achieve a bias of order $O\left(h^2\right)$ in Euclidean settings. The key lies in the specific form of the kernel function used in our estimator. Our smoothing term is constructed as $K_h\left(\delta\left(u^{\prime}, u\right)\right)=K_h\left(\left|u^{\prime}-u\right|\right)$, rather than the classical form $K_h\left(u^{\prime}-u\right)$. This difference is crucial: the integration of the first-order term $\int K_h\left(\left|u^{\prime}-u\right|\right)\left|u^{\prime}-u\right| d u^{\prime}$ does not vanish. Therefore, even under a Euclidean response space and apply a symmetric kernel, the first-order bias term in the Taylor expansion persists and dominates the convergence rate, leading to the $O(h)$ result. Therefore, the convergence rate established in Theorem \ref{mix_theorem1} is reasonable for a  general metric space $\mathcal{Y}$.

\begin{theorem}\label{mix_theorem2}
Under Assumptions (A1), (A2), (A6) and (A7)-(A9), we have
\begin{eqnarray*}
&&d( \widetilde{s}_{\oplus}( \bm x,u) ,\widehat{s}_\oplus(\bm x, u))  = O_p\big((n\varphi_{\mathcal{U},u}(h))^{\frac{-1}{2(\beta_2-1)}}\big),\\
\mbox{and} \, &&d( \widetilde{s}_\oplus(\bm x,u),\widehat{s}_\oplus(\bm x, u)) =O_{a.s.}\big(({n\varphi_{\mathcal{U},u}(h)}/{\log n})^{\frac{-1}{2(\beta_2-1)}}\big),
\end{eqnarray*}
where $\beta_2$ defined in Assumption (A6). Here, the almost sure convergence rate means that there exists a constant $M>0$, such that $$\mathbb{P}\big( \limsup_{n\to\infty} 
d(\widetilde{s}_{\oplus}(\bm x,u), \widehat{s}_{\oplus}(\bm x,u))({n\varphi_{\mathcal{U},u}(h)}/{\log n})^{\frac{1}{2(\beta_2-1)}}) 
\leq M\big)=1.$$
\end{theorem}
This theorem establishes the convergence rates of the stochastic term $d(\widetilde{s}_{\oplus}(\bm{x},u),\widehat{s}_{\oplus}(\bm{x},u))$. 
These rates depend on both the local curvature parameter $\beta_{2}$ in Assumption (A6) and the small ball probability ${\varphi}_{\mathcal U,u}(h)$. Essentially, the term $n\varphi_{\mathcal{U},u}(h)$ represents the effective local sample size, which directly determines the precision of the estimation in the neighborhood of $u$. It is worth noting that if $U$ lies in Euclidean space equipped with the Euclidean distance, then the small ball probability ${\varphi}_{\mathcal U,u}(h)=\int_{u-h}^{u+h}dF_\mathcal{U}$ is of order $h$.  Consequently, the stochastic convergence rate in Theorem \ref{mix_theorem2} reduces to $O_p\left((n h)^{-1 /2}\right)$, which aligns with the rate established in Theorem \ref{linear_theorem2}. Furthermore, the almost sure convergence rate is nearly optimal, differing from the convergence rate in probability only by a logarithmic factor.

\begin{corollary}
If the assumptions of Theorems~\ref{mix_theorem1} and \ref{mix_theorem2} hold, then
\begin{eqnarray*}
&&d( {{s}_{\oplus}({\bm{x},u}),{\widehat{s}}_{ \oplus  }( {\bm{x},u})}) = {O}_{p}\big(h^{\frac{1}{\beta_1 - 1}}+(n\varphi_{\mathcal{U},u}(h))^{\frac{-1}{2(\beta_2-1)}}\big),\\
 \mbox{and} \, &&d( {{s}_{\oplus}({\bm{x},u}),{\widehat{s}}_{ \oplus  }( {\bm{x},u})})= O_{a.s.}\big(h^{\frac{1}{\beta_1 - 1}}+(n\varphi_{\mathcal{U},u}(h)/\log n)^{\frac{-1}{2(\beta_2-1)}}\big).
\end{eqnarray*}
\end{corollary}
This corollary combines the bias term in Theorem~\ref{mix_theorem1} and the stochastic variation term in Theorem~\ref{mix_theorem2}. The convergence rate of the $d({{s}_{\oplus}(\bm{x},u),{\widehat{s}}_{\oplus}(\bm{x},u)})$ depends on the Lipschitz continuity of the conditional densities and the marginal density in Assumption (A10), and local geometry properties as quantified in (A5) and (A6) through $\beta_{1}$, $\beta_{2}$.

\section{Simulation studies}\label{sec:simulate}

In the simulation experiments, two types of responses are considered to demonstrate the performance of the varying-coefficient Fr\'echet regression model. The one is probability distributions and another one is symmetric positive definitive matrices. For every type of response, we present three examples where the predictor $U$ be chosen as the scalar, the density and the symmetric positive definite matrix. These examples cover both Euclidean predictor and non-Euclidean predictor. 

We denote the proposed varying-coefficient Fr\'echet regression as VFR. For comparison purposes, we also consider the following three competitors: (1) global Fr\'echet regression (GFR) proposed by \cite{PetersenFrechetregressionrandom2019}; (2) the local Fr\'echet regression (LFR) introduced by \cite{PetersenFrechetregressionrandom2019}; (3) the partially-global Fr\'echet regression (PFR) proposed by \cite{TuckerPartiallyglobalfrechetregression2025}. The computations for the GFR and the LFR can be carried out by R package \texttt{frechet} \citep{Chenfrechetstatisticalanalysis2020}. Following  \cite{PetersenFrechetregressionrandom2019}, a grid of bandwidths $\{0.05, 0.1, \dots, 0.3\}$ together with the Gaussian kernel are used for constructing LFR, PFR and VFR. The optimal bandwidth is selected by 10-fold cross-validation.

Based on the independent testing data \(\{{{{\widetilde{\mathbf{X}}}_{i},{\widetilde{U}}_{i},{\widetilde{Y}}_{i}}}\}^{\widetilde{n}}_{i=1}\) and a specific metric $d\br{\cdot,\cdot}$, we calculate a generalized mean squared error (GMSE) defined as $\widetilde{n}^{-1}\mathop{\sum }_{{i = 1}}^{\widetilde{n}}{{d}^{2}( {\widehat{s}}_{\oplus}( {\widetilde{\mathbf{X}}}_{i},{\widetilde{U}}_{i}),{\widetilde{Y}}_{i})}$, where ${\widehat{s}}_{\oplus}(\cdot,\cdot)$ is an estimated Fr\'echet regression function based on the training data using GFR, LFR, PFR and VFR, respectively. These criteria are used to evaluate the prediction performance of each method. In the following simulations, all the results are obtained by computing the averaged values of GMSE with 100 replications. In addition, training samples of size $n \in \{50,100,200\}$ and testing samples of size \(\widetilde{n}=1000\) were used for all examples. The raw predictors were generated in two steps:
(1) \({\left( {S}_{1},{S}_{2},{S}_{3},{S}_{4}\right) }^{\T}\) is multivariate Gaussian with \(\E\left( {S}_{j}\right)  = 0\) and \(\cov( {{S}_{j},{S}_{{j}^{\prime }}})  = {0.5 }^{\left| j - {j}^{\prime }\right| }\) for $1 \leq  j,{j}^{\prime } \leq  4$; (2) Set \(T = \Phi \left( {S}_{4}\right)\), where \(\Phi(\cdot)\) is the standard normal distribution function, so that $T \sim  Unif \left[{0,1}\right]$.

\subsection{Probability distribution as response}
Let \({\Omega }_{1}\) be the set of probability distributions. The 2-Wasserstein metric distance between two distributions with cumulative distribution functions \(H( \cdot )\) and \(G(\cdot )\) is defined as $$d_W(H,G)  = \left(\int_{0}^{1}{(H^{-1}(t)  - G^{-1}(t)) }^{2}dt\right)^{1/2}.$$

We denote \(\left( {{\Omega }_{1},{d}_{W}}\right)\) as the metric space of probability distributions equipped with the Wasserstein distance. Consider \(\mathcal{Y} \subseteq  {\Omega }_{1}\), the responses $Y$ represent the distribution function with the corresponding quantile function $Q\br{Y}$. Following \cite{PetersenFrechetregressionrandom2019}, for the sake of notational simplicity, we also denote the quantile function corresponding to $Y$ as $Y$. For computational tractability, we approximate the Wasserstein distance $d^2_W\br{H,G}$ using the discrete form $m^{-1}\sum^m_{i=1}\br{H^{-1}\br{t_i}-G^{-1}\br{t_i}}^2$, where $\{ t_1,\ldots,t_m\}$ denotes an equally spaced grid on the interval $[0,1]$ with $m = 20$ as in \cite{TuckerVariableselectionglobal2021}. 
\begin{example}\label{ex1}
(\(U\) is a scalar) In this example, we consider \(\mathcal{U} \subseteq  \mathbb{R}\). Set \({X}_{1} = {S}_{1},{X}_{2} = {S}_{2}, {X}_{3} = {S}_{3}\), and \(U = T\). The Fr\'echet regression function is given by
$$
{m}_{ \oplus  }\left( {\bm{x},u}\right)  = \E\left( {Y\left( \cdot \right)  \mid  \bm{X} = \bm{x},U = u}\right)  = \mu_0+\gamma_1u{x}_{1} + \gamma_2u^2{x}_{2} + \left( {{\sigma }_{0} + \gamma_3 \sin \left( {\pi u}\right)x_3 }\right) {\Phi }^{-1}\left( \cdot \right) .
$$
Conditional on \(\bm{X}\) and \(U\), the random response \(Y(\cdot)\) is generated by adding noise as follows: \(Y(\cdot)= \mu+\sigma {\Phi }^{-1}(\cdot)\), where \(\mu\mid\br{\bm{X},U} \sim  N( {\mu_0+\gamma_1U{X}_{1} + \gamma_2U^2{X}_{2},{\nu }_{1}})\) and \(\sigma  \mid  U \sim Gamma(\br{{\sigma }_{0} + \gamma_3 \sin ( {\pi U})X_3}^2/\nu_2,\nu_2/\br{{{\sigma }_{0} + \gamma_3 \sin( {\pi U})X_3 }})\) are independently sampled. The additional parameters are set to be \(\mu_0=1,\gamma_1=2, {\sigma }_{0} =2,{\nu }_{1} = 1\), \({\nu }_{2} = 2\) and $(\gamma_2,\gamma_3)\in \{ (1,1),(3,3)\}$ for two different Fr\'echet regression functions ${m}_{\oplus}\left( {\bm{x},u}\right)$.
\end{example}

\begin{example}\label{ex2}
(\(U\) is a density) In this example, we consider \(\mathcal{U} \subseteq  {\Omega }_{1}\). Set \({X}_{1} = {S}_{1},{X}_{2} = {S}_{2},{X}_{3} = {S}_{3}\). Further, we set $\E\left( {U\left( \cdot \right)  \mid  T = t}\right)  = {\mu }_{0u} + \left( {{\sigma }_{0u} + {\gamma }_{u}t}\right) {\Phi }^{-1}\left( \cdot \right).$
Conditional on \(T\), the predictor \(U(\cdot)\) is generated by adding noise as follows: \(U(\cdot) = {\mu }_{0u} + {\sigma }_{u}{\Phi }^{-1}(\cdot)\), where \({\sigma }_{u}\mid T \sim  Gamma( {{\left( {\sigma }_{0u} + {\gamma }_{u}T\right) }^{2}/{\nu }_{u},{\nu }_{u}/\left( {{\sigma }_{0u} + {\gamma }_{u}T}\right) })\) is independently sampled.
The Fr\'echet regression function is given by
$${m}_{ \oplus  }\left( {\bm{x},t}\right)  = \E\left( {Y\left(\cdot\right)  \mid  \bm{X} = \bm{x},T = t}\right)  = \mu_0+\gamma_1t{x}_{1} + \gamma_2t^2{x}_{2} + \left( {{\sigma }_{0} + \gamma_3 \sin \left( {\pi t}\right)x_3 }\right) {\Phi }^{-1}\left( \cdot \right).$$
Conditional on \(\bm{X}\) and \(T\), the random response \(Y(\cdot)\) is generated by adding noise as follows: \(Y(\cdot) = \mu  + \sigma {\Phi }^{-1}(\cdot)\), where $\mu  \mid  \br{\bm{X},T} \sim  N\left( { \mu_0+\gamma_1T{X}_{1} + \gamma_2T^2{X}_{2},{\nu }_{1}}\right)$ and $\sigma \mid T \sim Gamma(\br{{\sigma }_{0} + \gamma_3 \sin \left( {\pi T}\right)X_3}^2/\nu_2,\nu_2/\br{{{\sigma }_{0} + \gamma_3 \sin \left( {\pi T}\right) X_3}})$ are independently sampled. The additional parameters are set to be \(\mu_{0u}=0,\nu_u=1, \gamma_u=3, \sigma_{0u}= 2\). The other parameters are the same as in Example \ref{ex1}.
\end{example}

In the above Example, we can find that $Y$ depends on $U$ through the latent variable $T$. Further, because $U$ is no longer in  Euclidean space, we only can implement partially-global Fr\'echet regression model with local constant smoothing \citep{TuckerPartiallyglobalfrechetregression2025} and our varying-coefficient Fr\'echet regression model.

\begin{example}\label{ex3}
(\(U\) is a symmetric positive definite matrix) Consider \(\mathcal{U} \subseteq  {\Omega }_{2}\), which is the set of symmetric positive definite matrices. Let \({X}_{1} = {S}_{1}, {X}_{2} = {S}_{2}\) and \({X}_{3} = {S}_{3}\). Further, set
$\E(U \mid  T = t)  = \E{( A \mid T = t) }^{\T}\E( {A \mid  T = t}),$
where $\E(A \mid T = t)  = (\mu_{0u} + \beta_ut+\sigma_{0u} + \gamma_ut)  I + (\sigma_{0u} + \gamma_ut) V$, \(I\) denote an \(M \times M\) identity matrix and \(V = \left( {V}_{i,j}\right)\) denote an \(M \times  M\) matrix with \({V}_{i,j} = {I}_{\left( i<j\right) }\). Conditional on \(T\), the predictor \(U\) is generated by adding noise as follows: \(U = {A}^{\T}(N^{-1}\sum_{i=1}^NZ_iZ_i^\T)A\), where $(Z_1,\ldots,Z_N)$ are independently generated from $N_M(0,I_M)$,  $A = \br{{\mu }_{u}+{\sigma }_{u}}I + {\sigma }_{u}V$, ${\mu }_{u} = {\mu }_{0u} + {\beta }_{u}T$,  and ${\sigma }_{u} \mid  T$ is independently sampled  from $Gamma(( {\sigma }_{0u} + {\gamma }_{u}T)^{2}/{\nu }_{u},\nu_{u}/( \sigma_{0u} + {\gamma }_{u}T))$.
The Fr\'echet regression function is given by
$${m}_{ \oplus  }\left( {\bm{x},t}\right)  = \E\left( {Y\left( \cdot \right)  \mid  \bm{X} = \bm{x},T = t}\right)  = \mu_0+\gamma_1t{x}_{1} + \gamma_2t^2{x}_{2} + \left( {{\sigma }_{0} + \gamma_3 \sin \left( {\pi t}\right)x_3 }\right) {\Phi }^{-1}\left( \cdot \right).$$
The additional parameters are set to be \(M=2,N=10,\beta_u=1,\mu_{0u}=2\). The other settings are the same as in Example \ref{ex2}.
\end{example}

It can be seen that in the above Fr\'echet regression function ${m}_{ \oplus  }\left( {\bm{x},u}\right)$, parameter $\gamma_1$ governs the linear component, while $(\gamma_2,\gamma_3)$ determine the nonlinear part. Accordingly, we consider the setting $(\gamma_2,\gamma_3)\in\{(1,1),(3,3)\}$ to evaluate each method's prediction performance under different degrees of nonlinearity. As shown in Table \ref{simulation1}, GFR remains competitive when the degree of nonlinearity is relatively low. The advantage of VFR  over  GFR  is more pronounced in the higher nonlinearity case $(\gamma_2,\gamma_3)=(3,3)$.
In Example \ref{ex1}, it is seen that the proposed VFR is significantly superior to GFR, LFR and PFR when $U$ is a scalar predictor. The VFR also performs well in both Example \ref{ex2} and Example \ref{ex3} when the predictor $U$ is non-Euclidean. We also observe that the GMSE of VFR decreases as the training sample size $n$ increases.

\subsection{Symmetric positive definite matrix as response}
Let \({\Omega}_{2}\) be the set of symmetric positive definite matrices. The Cholesky decomposition metric distance between two symmetric positive definite matrices \({{P}}_{1}\) and \({{P}}_{2}\) is defined as
\[
{d}_{C}( {{{P}}_{1},{{P}}_{2}})  =(\text{trace}( {{( {{P}}_{1}^{1/2} - {{P}}_{2}^{1/2}) }^{\T}( {{{P}}_{1}^{1/2} - {{P}}_{2}^{1/2}}) }))^{1/2}.
\]
We denote \(({{\Omega }_{2},{d}_{C}})\) as the metric space of symmetric positive definite matrices equipped with the Cholesky decomposition distance.  
Consider \(\mathcal{Y} \subseteq {\Omega}_{2}\). Let \({I}\) denotes an \(M \times  M\) identity matrix and \({V} = \left( {V}_{i,j}\right)\) denotes an \(M \times  M\) matrix where \({V}_{i,j} = {I}_{\left( i<j \right) }\).

\begin{example}\label{ex4}
(\(U\) is a scalar) Consider \(\mathcal{U} \subseteq  \mathbb{R}\). We set \({X}_{1} = {S}_{1},{X}_{2} = {S}_{2},{X}_{3} = {S}_{3}\), and \(U = T\). The Fr\'echet regression function is given by ${m}_{ \oplus  }\left( {\bm{x},u}\right)  = \E\left( {Y \mid  \bm{X} = \bm{x}, U=u}\right)  = \E{\left( B \mid  \bm{X} = \bm{x},U = u\right) }^{\T} \E\left( {B \mid  \bm{X} = \bm{x},U = u}\right)$,
where $\E({B \mid \bm{X} = \bm{x},U = u}) = (\mu_0+\gamma_1 u{x}_{1} + \gamma_2u^2{x}_{2}+( {{\sigma }_{0} + \gamma_3 \sin ( {\pi u})x_3 })) {I} + ({\sigma }_{0} + \gamma_3 \sin ( {\pi u})x_3)V$.
Conditional on \(\bm{X}\) and \(U\), the random response \(Y\) is generated by adding noise as follows: \(Y = {B}^{\T}(N^{-1}\sum_{i=1}^NZ_iZ_i^\T)B\), where \(B = \left( {\mu  + \sigma }\right){I} +  \sigma V\), $\mu \mid (\bm{X},U) \sim  N \left( \mu_0+\gamma_1UX_1+\gamma_2U^2X_2,{\nu }_{1}\right)$, and \(\sigma  \mid  (\bm{X},U) \sim {Gamma}({{\br{{\sigma }_{0} + \gamma_3 \sin \left( {\pi U}\right)X_3} }^{2}/{\nu }_{2},{\nu }_{2}/\br{{\sigma }_{0} + \gamma_3 \sin \left( {\pi U}\right)X_3} })\) are independently sampled. The additional parameters are the same as in Example \ref{ex3}.    
\end{example}
\begin{example}\label{ex5}
(\(U\) is a density) Here we consider \(\mathcal{U} \subseteq  {\Omega }_{1}\). Set \({X}_{1} = {S}_{1},{X}_{2} = {S}_{2}\) and \({X}_{3} = {S}_{3}\). Further, 
\[
\E\left( {U\left( \cdot \right)  \mid  T = t}\right)  = {\mu }_{0u} + \left( {{\sigma }_{0u} + {\gamma }_{u}t}\right) {\Phi }^{-1}\left( \cdot \right) .
\]
Conditional on \(T\), the predictor \(U(\cdot)\) is generated by adding noise as follows: \(U(\cdot) = {\mu }_{0u} + {\sigma }_{u}{\Phi }^{-1}(\cdot)\) with \({\sigma }_{u} \mid  T \sim  Gamma( {{\left( {\sigma }_{0u} + {\gamma }_{u}T\right) }^{2}/{\nu }_{u},{\nu }_{u}/\left( {{\sigma }_{0u} + {\gamma }_{u}T}\right) })\) being independently sampled.
The Fr\'echet regression function is given by ${m}_{ \oplus  }\left( {\bm{x},t}\right)  = \E\left( {Y \mid  \bm{X} = \bm{x},T = t}\right)  = \E{\left( B \mid  \bm{X} = \bm{x},T = t\right) }^{\T}\E\left( {B \mid  \bm{X} = \bm{x},T = t}\right)$, where $\E({B \mid \bm{X} = \bm{x},T = t})=(\mu_0+\gamma_1 t{x}_{1} + \gamma_2t^2{x}_{2}+({{\sigma }_{0} + \gamma_3 \sin ( {\pi t})x_3 })) I +({\sigma }_{0} + \gamma_3 \sin(\pi t)x_3) V$.
Conditional on \(\bm{X}\) and \(T\), the random response \(Y\) is generated by adding noise as follows: \(Y = {B}^{\T}(N^{-1}\sum_{i=1}^NZ_iZ_i^\T)B\), where \(B = (\mu  + \sigma){I} + \sigma V\), $\mu \mid (\bm{X},T) \sim  N ( \mu_0+\gamma_1 T{X}_{1} + \gamma_2T^2{X}_{2},{\nu }_{1})$, and \(\sigma  \mid  (\bm{X},T) \sim  {Gamma}( {{\br{{\sigma }_{0} + \gamma_3 \sin ( {\pi T})X_3} }^{2}/{\nu }_{2},{\nu }_{2}/\br{{\sigma }_{0} + \gamma_3 \sin ( {\pi T})X_3}})\) are independently sampled. The additional parameters are the same as in Example \ref{ex3}. 
\end{example}
\begin{example}\label{ex6}
(\(U\) is a symmetric positive definite matrix) We consider \(\mathcal{U} \subseteq  {\Omega }_{2}\). Set \({X}_{1} = {S}_{1},{X}_{2} = {S}_{2}\) and \({X}_{3} = {S}_{3}\). Further, we set
$$
\E(U \mid  T = t)  = \E( A \mid  T = t)^\T\,\E(A \mid T = t),
$$
where $\E(A \mid  T = t)  = (\mu_{0u} + \beta_{u}t  + \sigma_{0u} + {\gamma }_{u}t)  {I}_{M_u} +(\sigma_{0u} + \gamma_ut) U_{M_u}$. Conditional on \(T\), the predictor \(U\) is generated by adding perturbation as follows: \(U = {A}^{\T}(N^{-1}\sum_{i=1}^NZ_iZ_i^\T)A\), where $(Z_1,\ldots,Z_N)$ are independently generated from $N_M(0,I_M)$, $A = ( {{\mu }_{u} + {\sigma }_{u}}) {I}_{{M}_{u}} + {\sigma }_{u}{U}_{{M}_{u}}$, $\mu_{u} \mid  T \sim N( \mu_{0u}+\beta_{u}T,\nu_{u1})$, and $\sigma_u\mid T\sim Gamma({(\sigma_{0u} + \gamma_{u}T)}^{2}/\nu_{u2},\nu_{u2}/(\sigma_{0u} + \gamma_{u}T))$. The Fr\'echet regression function is given by
$m_\oplus( {\bm{x},t}) =\E( {Y \mid  \bm{X} = \bm{x},T = t})  = \E{( B \mid  \bm{X} = \bm{x},T = t) }^\T\E( {B \mid \bm{X} = \bm{x},T = t})$, which is the same as in Example~\ref{ex5}.
Conditional on \(\bm{X}\) and $T$, the random response $Y$ is generated by adding noise as in Example~\ref{ex5}. The additional parameters are set as $\beta_u=\gamma_u=4, M_u=2, \nu_{u1}=\nu_{u2}=1$. The other parameters are the same as in Example \ref{ex5}.   
\end{example}

Table \ref{simulation2} reports the results when the response is symmetric positive definite matrix under different degrees of nonlinearity setting. It can be seen that the superiority of VFR is more pronounced than the remaining three models under higher nonlinearity case $(\gamma_2,\gamma_3)=(3,3)$. Moreover, VFR outperforms other competitive Fr\'echet regression methods when $U$ is the Euclidean predictor in Example \ref{ex4}, and achieves lower GMSE than PFR when the predictor $U$ comes from the non-Euclidean space in Example \ref{ex5} and \ref{ex6}.

\begin{table}[htbp]
\captionsetup{font={stretch=1.0}}
\centering
\renewcommand{\arraystretch}{0.65}
\caption{The averaged GMSE of various methods and the associated standard errors (in parenthesis) when the response is probability distributions.}
\label{simulation1}
\begin{threeparttable} 
\setlength{\tabcolsep}{0pt}
\begin{tabular}{c@{\hskip 15pt}c@{\hskip 15pt}c@{\hskip 3pt}c@{\hskip 3pt}c@{\hskip 3pt}c@{\hskip 6pt}c@{\hskip 3pt}c@{\hskip 3pt}c@{\hskip 3pt}c}
\toprule
\multicolumn{2}{c}{} & \multicolumn{4}{c}{$(\gamma_2,\gamma_3)=(1,1)$} & \multicolumn{4}{c}{$(\gamma_2,\gamma_3)=(3,3)$} \\
\cmidrule(lr){3-6} \cmidrule(lr){7-10}
 & & GFR & LFR & PFR & VFR & GFR & LFR & PFR & VFR \\
\toprule
\multirow{5}{*}{Example \ref{ex1}} & \multirow{1.6}{*}{$n=50$} & 3.548 & 273.609 & 3.818 & \bf{3.367} & 5.109 & 68.985 & 5.962 & \bf{4.516} \\[-0.4em]
 & &(0.027) & (218.377) & (0.047) & (0.027) & (0.053) & (15.468) & (0.096)&(0.056) \\

 & \multirow{1.6}{*}{$n = 100$} & 3.334 & 29.809 & 3.690 & \bf{3.142} &4.984 & 214.133 & 5.879 & \bf{4.301} \\[-0.4em]
 & & (0.017) & (12.290) & (0.042) & (0.017) & (0.082) & (110.749) & (0.119) & (0.085) \\

 & \multirow{1.6}{*}{$n = 200$} & 3.226 & 9.696 & 3.517 & \bf{2.985} & 4.696 & 13.547 & 5.562 & \bf{3.935} \\[-0.4em]
 & & (0.015) & (0.454) & (0.040) & (0.015) & (0.041) & (2.256) & (0.085) & (0.039) \\
\midrule
\multirow{5}{*}{Example \ref{ex2}} & \multirow{1.6}{*}{$n = 50$} & \multirow{1.6}{*}{--} & \multirow{1.6}{*}{--} & 3.879 & \bf{3.698} & \multirow{1.6}{*}{--} & \multirow{1.6}{*}{--} & 5.576 & \bf{5.248} \\[-0.4em]
 & &  &  & (0.047) & (0.033) &  &  & (0.070) & (0.048) \\

 & \multirow{1.6}{*}{$n = 100$} & \multirow{1.6}{*}{--} & \multirow{1.6}{*}{--} & 3.503 & \bf{3.432} & \multirow{1.6}{*}{--} & \multirow{1.6}{*}{--} & 4.981 & \bf{4.838} \\[-0.4em]
 & &  & & (0.027) & (0.025) &  &  & (0.036) & (0.034) \\

 & \multirow{1.6}{*}{$n = 200$} & \multirow{1.6}{*}{--} & \multirow{1.6}{*}{--} & 3.349 & \bf{3.272} & \multirow{1.6}{*}{--} & \multirow{1.6}{*}{--} & 4.782 & \bf{4.611} \\[-0.4em]
 & &  &  & (0.018) & (0.017) &  & & (0.031) & (0.027) \\
\midrule
\multirow{5}{*}{Example \ref{ex3}} & \multirow{1.6}{*}{$n = 50$} & \multirow{1.6}{*}{--} & \multirow{1.6}{*}{--} & 5.853 & \bf{4.105} & \multirow{1.6}{*}{--} & \multirow{1.6}{*}{--} & 10.244 & \bf{5.913} \\[-0.4em]
 & &  &  & (0.190) & (0.062) &  &  & (0.593) & (0.092) \\

 & \multirow{1.6}{*}{$n = 100$} & \multirow{1.6}{*}{--} & \multirow{1.6}{*}{--} & 5.816 & \bf{4.084} & \multirow{1.6}{*}{--} & \multirow{1.6}{*}{--} & 9.322 & \bf{5.590} \\[-0.4em]
 & &  &  & (0.235) & (0.108) &  &  & (0.330) & (0.114) \\

 & \multirow{1.6}{*}{$n = 200$} & \multirow{1.6}{*}{--} & \multirow{1.6}{*}{--} & 5.354 & \bf{3.662} & \multirow{1.6}{*}{--} & \multirow{1.6}{*}{--} & 9.104 & \bf{5.529} \\[-0.4em]
 & & &  & (0.191) & (0.035) &  &  & (0.299) & (0.091) \\
\bottomrule
\end{tabular}
\begin{tablenotes}
\footnotesize
\item Note: ``--” denotes that the corresponding method cannot be applied in these examples.
\end{tablenotes}
\end{threeparttable} 
\end{table}

\begin{table}[htbp]
\captionsetup{font={stretch=1.0}}
\centering
\renewcommand{\arraystretch}{0.65}
\caption{The averaged GMSE of various methods and the associated standard errors (in parenthesis) when the response is symmetric positive definite matrices.}
\label{simulation2}
\begin{threeparttable} 
\setlength{\tabcolsep}{0pt}
\begin{tabular}{c@{\hskip 13pt}c@{\hskip 13pt}c@{\hskip 3pt}c@{\hskip 3pt}c@{\hskip 3pt}c@{\hskip 6pt}c@{\hskip 4pt}c@{\hskip 4pt}c@{\hskip 4pt}c}
\toprule
\multicolumn{2}{c}{} & \multicolumn{4}{c}{$(\gamma_2,\gamma_3)=(1,1)$} & \multicolumn{4}{c}{$(\gamma_2,\gamma_3)=(3,3)$} \\
\cmidrule(lr){3-6} \cmidrule(lr){7-10}
 & & GFR & LFR & PFR & VFR & GFR & LFR & PFR & VFR \\
\toprule
\multirow{5}{*}{Example \ref{ex4}} & \multirow{1.6}{*}{$n = 50$} & 15.971 & 399.418 & 17.270 & \bf{15.781} & 26.524 & 1190.453 & 26.406 & \bf{21.734} \\[-0.4em]
 & & (0.737) & (109.056) & (0.758) & (0.737) & (1.337) & (586.919) & (1.120) & (1.032) \\

 & \multirow{1.6}{*}{$n = 100$} & 15.923 & 90.522 & 16.420 & \bf{15.244} & 25.288 & 87.716 & 24.617 & \bf{21.629} \\[-0.4em]
 & & (0.876) & (23.212) & (0.846) & (0.845) & (1.250) & (11.597) & (1.106) & (1.064) \\

 & \multirow{1.6}{*}{$n = 200$} & 16.138 & 47.509 & 16.653 & \bf{15.482} & 21.426 & 44.554 & 22.106 & \bf{18.758} \\[-0.4em]
 & & (0.745) & (5.716) & (0.720) & (0.700) & (1.011) & (3.874) & (0.840) & (0.811) \\
\midrule
\multirow{5}{*}{Example \ref{ex5}} & \multirow{1.6}{*}{$n = 50$} & \multirow{1.6}{*}{--} & \multirow{1.6}{*}{--} & 15.854 & \bf{15.697} & \multirow{1.6}{*}{--} & \multirow{1.6}{*}{--} & 23.516 & \bf{22.620} \\[-0.4em]
 & &  &  & (0.628) & (0.634) &  &  & (1.143) & (1.127) \\

 & \multirow{1.6}{*}{$n = 100$} & \multirow{1.6}{*}{--} & \multirow{1.6}{*}{--} & 15.964 & \bf{15.846} & \multirow{1.6}{*}{--} & \multirow{1.6}{*}{--} & 22.235 & \bf{21.948} \\[-0.4em]
 & &  &  & (0.657 & (0.653) &  &  & (1.297) & (1.315) \\

 & \multirow{1.6}{*}{$n = 200$} & \multirow{1.6}{*}{--} & \multirow{1.6}{*}{--} & 15.154 & \bf{15.032} & \multirow{1.6}{*}{--} & \multirow{1.6}{*}{--} & 21.966 & \bf{21.594} \\[-0.4em]
 & &  & & (0.665) & (0.669) &  &  & (0.992) & (0.971) \\
\midrule
\multirow{5}{*}{Example \ref{ex6}} & \multirow{1.6}{*}{$n = 50$} & \multirow{1.6}{*}{--} & \multirow{1.6}{*}{--} & 22.669 & \bf{17.297} & \multirow{1.6}{*}{--} & \multirow{1.6}{*}{--} & 35.650 & \bf{24.988} \\[-0.4em]
 & &  &  & (1.263) & (0.779) &  &  & (1.509) & (1.060) \\

 & \multirow{1.6}{*}{$n = 100$} & \multirow{1.6}{*}{--} & \multirow{1.6}{*}{--} & 24.284 & \bf{18.474} & \multirow{1.6}{*}{--} & \multirow{1.6}{*}{--} & 37.313 & \bf{25.259} \\[-0.4em]
 & &  &  & (1.057) & (0.868) &  &  & (1.716) & (1.102) \\

 & \multirow{1.6}{*}{$n = 200$} & \multirow{1.6}{*}{--} & \multirow{1.6}{*}{--} & 26.526 & \bf{19.399} & \multirow{1.6}{*}{--} & \multirow{1.6}{*}{--} & 38.731 & \bf{26.273} \\[-0.4em]
 & &  &  & (1.871) & (0.938) &  &  & (2.279) & (1.311) \\
\bottomrule
\end{tabular}
\begin{tablenotes}
\footnotesize
\item Note: ``--” denotes that the corresponding method cannot be applied in these examples.
\end{tablenotes}
\end{threeparttable}
\end{table}

\section{Real data analysis}\label{sec:apply}

In this section, we apply the proposed varying-coefficient Fr\'echet regression and other version of Fr\'echet regression to the human mortality dataset. The goal is to model the dependence of age-at-death distribution in 2013, treated as random objects, on country-specific covariates. Compared with traditional summary measures such as the death rate, modeling the age-at-death distribution as a random object provides deeper insights into human longevity.

For this purposes, we consider the random object response derived from \cite{HMD} (HMD, \url{http://www.mortality.org}), which contains life tables for 39 countries in 2013. To focus on adult mortality, we restrict our analysis to histograms over the age range [20,100]. As for the predictors, extensive research has documented the effects of socioeconomic, environmental and other related factors on health outcomes. Motivated by the literature, we select five predictors that are potentially relevant to explaining cross-country mortality patterns. These include year-on-year percentage change in GDP (GDPC), carbon dioxide emissions in metric tons per capita (CO2E), current health care expenditure as a percentage of GDP (HCE), the human development index (HDI), and infant mortality per 1000 live births (IM). The CO2E is used as the local predictor $U$ in VFR. Hence, $\bm{X}_i\in\mathbb{R}^4$ and $U_i\in \mathbb{R}$ constitute the predictors for the $i$-th country, $i = 1, ..., 39$. The data are collected by \cite{BhattacharjeeSingleindexfrechet2023}.

To evaluate the prediction performance of the proposed method, we randomly select $n_{\text{train}}$ observations as the training data and use the remaining $n_{\text{test}}$ observations as the test data. Then, GFR, LFR, PFR and VFR are accessed by the mean squared prediction errors: $\text{MSPE}=n_{\text{test}}^{-1}\sum_{i=1}^{n_{\text{test}}}d^2_W(Y_i,\widehat{Y}_i)$. Here, $Y_i$ is the $i$-th testing observation and $\widehat{Y}_i$ represents the prediction for each method based on the training data. We repeat the above procedure 50 times and set the training sample size as $n_{\text{train}}\in\{20,25,30\}$.


\begin{table}[htbp]
\captionsetup{font={stretch=1.0}}
\centering
\renewcommand{\arraystretch}{0.65}
\caption{The MSPE values (with standard errors in parentheses) of each methods.}
\label{tab:mse_results}
\begin{tabular}{cccc}
\toprule
Methods & $n_{\text{train}}=20$ & $n_{\text{train}}=25$ & $n_{\text{train}}=30$ \\
\midrule
GFR & 7.203 (0.368) & 6.353 (0.343) & 5.775 (0.375) \\
LFR & 53.005 (8.542) & 83.856 (53.243) & 23.307 (1.727) \\
PFR  & 10.366 (0.705) & 10.367 (0.754) & 9.537 (0.763) \\
VFR  & \textbf{6.944} (0.357) & \textbf{6.161} (0.347) & \textbf{5.562} (0.390) \\
\bottomrule
\end{tabular}
\end{table}

Table \ref{tab:mse_results} reports the mean MSPEs and standard errors for various Fr\'echet regression. Among the five candidate predictors considered as the local predictor $U$ in VFR, CO2E yielded the best performance in terms of predictive accuracy MSPE. This result is reasonable because CO2E emissions exhibit a clear long-term trend and are closely tied to environmental and health risks that shape the age-at-death distribution \citep{AzimiUnveilinghealthconsequences2024}. Moreover, VFR always achieves smaller mean MSPE values than GFR, LFR, and PFR across all scenarios, indicating that our model provides superior predictive performance for the human mortality data.

\section{Discussion}\label{sec:discuss}
In this paper, we introduce a varying-coefficient Fr\'echet regression model and its corresponding estimators to handle both Euclidean and Non-Euclidean predictors. Additionally, we derive the convergence rates of the proposed estimators. Several examples with random objects are used to show the performance of varying-coefficient Fr\'echet regression model. Taken together, our results provide a foundation for further methodological and theoretical developments in modeling non-Euclidean data.

Despite the progress made in this work, a number of challenges and opportunities for further development remain. The incorporation of variable selection methods into Fr\'echet regression analysis would greatly enhance its applicability. A unified approach to variable selection for varying-coefficient models has been proposed by \cite{Tangunifiedvariableselection2012}, and it is natural to consider extending this framework to non-Euclidean responses. However, in the absence of explicit coefficients in the VFR model, it remains unclear how existing methods can be adapted to this more complex setting. Another important direction is the development of inference tools to understand the relationships between random object response and predictor. For the broader applicability of Fr\'echet regression, it is essential to clarify the significance of predictor effects. \cite{DubeyFrechetanalysisvariance2019} introduced a Fr\'echet analysis of variance for random objects, including a test statistic and its asymptotic distribution. A natural direction for future work is to investigate whether likelihood ratio tests can be developed for general metric spaces, thereby offering a unified framework for assessing and testing variable significance in Fr\'echet regression.

\clearpage

\clearpage
\bibliographystyle{agsm}
\bibliography{reference}

\end{document}